\documentclass[11pt]{article}  
  
\begin{document}

\title{Nonequilibrium phase transitions in a model for the origin of life}

\author{Claudia P. Ferreira and 
J. F. Fontanari \\
 Instituto de F\'{\i}sica de S\~ao Carlos \\
 Universidade de S\~ao Paulo \\
 Caixa Postal 369 \\
 13560-970 S\~ao Carlos SP, Brazil 
}

\date{}

\maketitle

\begin{abstract}
The  requisites  for the
persistence of small colonies of self-replicating molecules
living in a two-dimensional lattice are investigated analytically
in the infinite diffusion or mean-field limit and through Monte Carlo
simulations
in the position-fixed or  contact process limit. The molecules
are modeled by hypercyclic replicators $A$ which are capable of
replicating via binary fission $A + E \to 2A$  with production rate $s$ as well
as via catalytically assisted replication $2A + E \to 3A$
with rate $c$. In addition, a molecule can degrade into its source materials $E$ with
rate $\gamma$. In the
asymptotic regime  the system can be characterized by the
presence (active phase) and the absence (empty phase) of replicators in the
lattice. In both diffusion regimes, we find that
for small values of the ratio $c/\gamma$  these phases
are separated by a second-order 
phase transition  which is in the universality class of the directed percolation,
while for small values of $s/\gamma$ the phase transition is of first order.
Furthermore, we show the suitability of the dynamic Monte Carlo method,
which  is
based on  the analysis of the spreading behavior of a few
active cells in the center of an otherwise infinite empty lattice,
to address the problem of the emergence of replicators. 
Rather surprisingly, we show that this method allows an unambiguous 
identification of the order of the nonequilibrium phase transition. 
\end{abstract}

\bigskip


\newpage

\small 

%--------------------------------------------

\section{Introduction}\label{sec:level1}

%--------------------------------------------

The most fundamental event in the history of life was probably
the appearance via spontaneous creation
of a molecule  capable of replicating itself (replicator). 
Given a possible mechanism
of replication, which in this case is some form of template activity, 
the evolution of such replicators has been extensively
investigated through the chemical kinetics formalism
put forward by Eigen and co-workers in the seventies \cite{Eigen,hyper}.
Those studies have raised a series of objections to the simplistic
view of the emergence of a complex organism from a collection
of competing  species of replicators. For instance, the finding that
the length of a molecule (polynucleotide) is limited due to the 
finite replication accuracy per nucleotide has prompted the proposal
of models that incorporated some sort of cooperation between
the replicators, such as the molecular catalytic feedback networks
termed hypercycles \cite{hyper}. These models, however, have attracted
their own criticisms since, as pointed out by Maynard Smith, giving catalytic
support in such molecular  networks is in fact an altruistic behavior 
and so they are extremely vulnerable to the presence of parasites, i.e.,
molecules that do not reciprocate the  catalytic support they receive
\cite{Maynard}. A possible solution to this problem is provided by
the structured deme formulation of group selection \cite{Wilson}, where
it is  assumed that the
replicators are spatially localized, say, in rock crevices or water droplets,
so that the benefits accrued from  cooperation are directed mostly to the
members of the catalytic network \cite{Michod,SM,Alves}. Yet another 
successful approach
to the problem of resistance against parasites is based on 
a reaction-diffusion system where replication and diffusion
taking place on an adsorbing surface generate
self-organized  spiral structures
\cite{Boerlijst1,Boerlijst2}. Interestingly, as these
spatial structures, which greatly increase the stability of the
hypercycles
against parasites,  can be viewed as super-organisms that approach
is also related to the
group selection theory \cite{Boerlijst1}.

Since in the prebiotic or  chemical evolution  context, natural selection 
is essentially the dynamics of replicators, it is not surprising
that most of the studies in this subject have focused almost exclusively
on the competition between  replicators, among which
the so-called malthusian and hypercyclic replicators are the most
important \cite{hyper,Michod}. The former corresponds to the simplest
reproduction process, namely, the binary fission of a parent replicator 
and is modelled by the chemical reaction 
\begin{equation}\label{reac_malt}
A + E \stackrel{s}{\rightarrow} 2A     
\end{equation}
where $A$ is the replicator and $E$ is the source materials
(mononucleotide resources). It is well-known
that the concentration of $A$ grows exponentially with the rate constant
$s$, provided that the concentration  of $E$ is kept constant, hence the
name malthusian replicator. To avoid this explosive growth, one usually imposes
a constraint on the total concentration of replicators which can
be implemented in practice by a dilution flux \cite{Eigen}. Alternatively,
one can allow the replicators to be degraded by hydrolysis into its 
mononucleotide components $E$ according to
the reaction
\begin{equation}\label{reac_decay}
A \stackrel{\gamma}{\rightarrow} E     
\end{equation}
which seems a more natural approach to limit the growth of $A$.

As best exemplified by sexual reproduction, there are situations
that cannot be described by (\ref{reac_malt}) since
two replicators are necessary to produce a third one.
In this case  the corresponding chemical reaction is
\begin{equation}\label{reac_hyper}
2A + E \stackrel{c}{\rightarrow} 3A     
\end{equation}
which leads to a hyperbolic growth of the concentration of  replicator $A$ 
\cite{hyper}.
A hypercyclic replicator is defined as one that can replicate itself using 
both (\ref{reac_malt}) and (\ref{reac_hyper}) reaction
schemes. Actually, the term hypercycle derives from the superimposition
of the catalytic replication cycle (\ref{reac_hyper}) on the
self-replication cycle (\ref{reac_malt}).
Of course, the limit $c=0$ corresponds to the malthusian replicator
while $s=0$ can be associated to an obligatory sexual replicator.

In contrast to previous works that have concentrated on the competition between 
replicators either of the same kind but with different
production rates \cite{Eigen,hyper} or of different kinds \cite{Michod},
in this paper we address a more fundamental problem that has received
comparatively little attention, namely, the stability of the different
kinds of replicators, viewed here as an active (ordered) phase of the molecular
system, against the empty (disordered) phase composed of the resource
materials only. This lack of interest was probably due to the fact that the
usual kinetics formalism used to study the dynamics of
replicators does not represent the mononucleotide resource dynamics 
explicitly (see however \cite{Chacon,Cronhjort}), thus precluding  the 
study of the issues  addressed in the present contribution. More pointedly,
we consider the dynamics of a population of identical hypercyclic replicators on
a lattice space both in the deterministic infinite diffusion (mean-field) limit and 
in the stochastic position-fixed (contact process) limit  where each
replicator on a lattice cell  never moves. The last limit  is particularly
interesting because it allows the connection between the replicator models
and some standard models of nonequilibrium phase transition in a lattice
(e.g. directed percolation) \cite{Grass-AP,Grass-MB,Grass-JPA}. As a result, 
the powerful
analytical tools of  statistical mechanics
can be readily used to advance our understanding of the evolution of replicators.
Of particular relevance is the so-called dynamic Monte Carlo method 
whose idea is to set
the system initially in the empty state with a seed of replicators in the
center of the lattice and then study the subsequent spreading of activity
\cite{Grass-AP,Grass-MB,Grass-JPA}. More importantly, the thorough analysis
of both limits exposes the limitations of the  widely used deterministic chemical
kinetics or mean-field formalism to study the problem of the emergence of life.

The remainder of the paper is organized as follows. In Sec.\ \ref{sec:level2}
we present 
the set of rules that govern the evolution of a population of
hypercyclic replicators in a two-dimensional square lattice. The mean-field or 
infinite diffusion limit, which models an ideally mixed medium, is studied
analytically in Sec.\ \ref{sec:level3}. The results are summarized in a phase 
diagram   showing the regions of stability of
the empty and active regimes in the space of the control parameters of the model. 
Those regions are delimitated by 
continuous as well as discontinuous transition lines that
join at a tricritical point. In Sec.\ \ref{sec:level4} we study the
position-fixed or contact process limit using mainly the dynamic Monte
Carlo method which allows the computation of the  critical dynamic 
exponents that describe quantitatively the spreading of a vanishingly
small population of replicators.
Finally, some concluding remarks are presented in
Sec.\ \ref{sec:level5}. In particular, we compare the hypercyclic replicator
model with Schl\"ogl's 
models of nonequilibrium phase transition in reaction-diffusion systems
\cite{Schlogl}.

%--------------------------------------------

\section{The model}\label{sec:level2}

%--------------------------------------------

It has been suggested that chemical
evolution started with a surface-bonded autocatalytic chemical network
as there are enormous thermodynamics
and kinetics advantages of surface binding reactions, especially
in the case of reactions that require unlikely collisions of many reactants
\cite{Wach}. The binding must be strong enough  to keep the reactants
on the surface but also flexible enough to allow their slow migration on it.
Interestingly, if this proposal proves correct it will probably lead to 
the replacement of the popular notion of a primitive soup by that of a 
primitive pizza instead \cite{book}.  
Accordingly, we define our replicator model in a two-dimensional 
space consisting of $L \times L$ cells in a
square toroidal lattice. 
Each cell is either empty or occupied by a 
replicator and it is assumed that an empty cell contains all source material 
required to assemble a new replicator.
The evolution of the population of replicators is governed
by the following local rules:
\begin{itemize}
\item[(1)] A replicator has a probability $\gamma$ of decaying;
after decay the cell becomes empty. This rule is motivated by the hydrolysis 
reaction (\ref{reac_decay}).
\item[(2)] A replicator in one of the four first neighbor cells (von
Neumann neighborhood) of an empty cell can replicate into that cell
with probability $s$. This process is referred to as non-catalised
self-replication and is motivated by the reaction (\ref{reac_malt}). 
\item[(3)] Regardless of the previous rule, a replicator in the von
Neumann neighborhood of an empty cell can replicate into  that cell
if there are other replicators in the intersection of the Moore neighborhoods  
of both cells. The probability of this type of replication, which is motivated
by reaction (\ref{reac_hyper}), is 
$c$ for each pair of replicators. We recall that the Moore neighborhod of
a given cell consists of its  first and second nearest 
neighbors, adding up to eight cells.
\end{itemize}
Hence in the extreme situation where an empty cell is surrounded by eight replicators,
it can become occupied with probability $4 s + 16 c$. To carry out the simulations
we choose the parameters $c$ and $s$ such that $4 s + 16 c \leq 1$. 
These rules are applied simultaneously to all cells in the lattice so
our model can be viewed as a two-dimensional stochastic cellular automaton.
Actually the model is essentially an adaptation to one-membered hypercycles of 
the spatial cellular automaton model of multi-membered hypercycles proposed by
Boerlijst and Hogeweg   \cite{Boerlijst1}. The dynamics defined by the 
rules given above  is manifestly  irreversible and, in particular, the
state characterized by empty cells only is an absorbing state, i.e., a configuration
from which the system cannot escape. In this sense, the principle of detailed
balance is broken and the active stationary state is in fact in nonequilibrium.
Although the more realistic situation is a diffusion-controled reaction 
where each reactant can move randomly on the lattice, in this paper
we choose to study in detail the simpler extreme cases
of infinite diffusion (mean field) and no diffusion (contact process).
Of course, we hope that features common to both limits will be present
in the finite diffusion situation as well.

%----------------------------------------------------------------
%
\section{The mean-field limit}\label{sec:level3}
%
%----------------------------------------------------------------

\begin{figure}[tbp]
\vspace{6.5cm}
\includegraphics{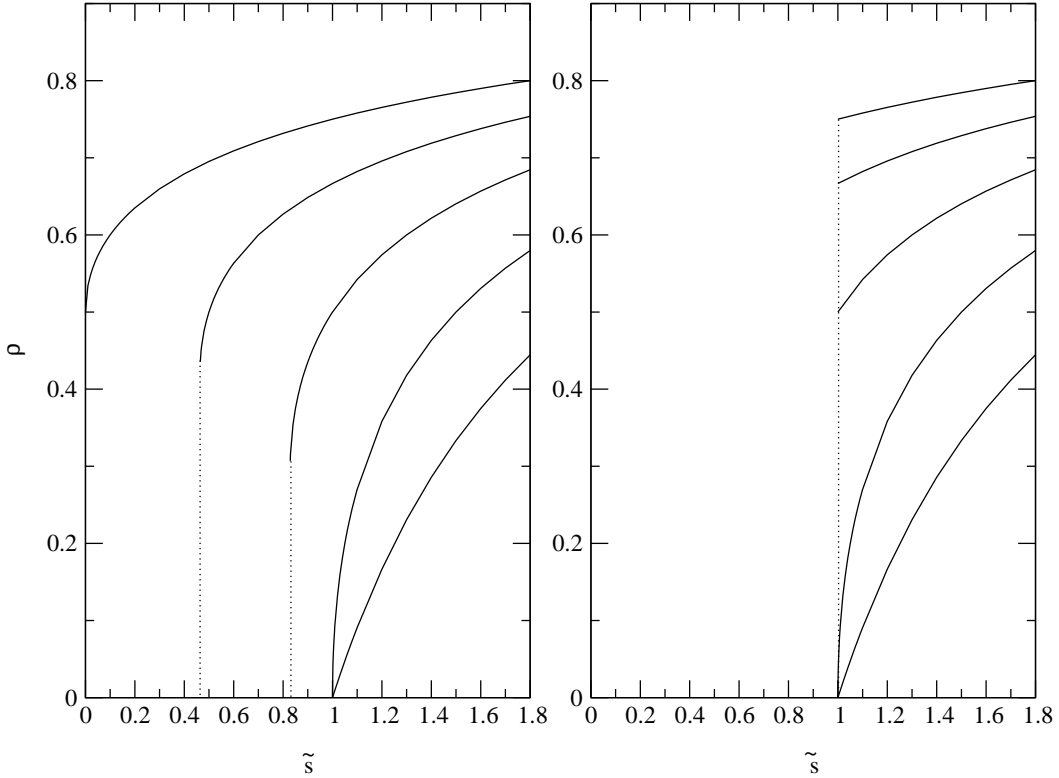} 
\vspace{2cm}
\caption{Mean-field steady-state concentration of active sites as a function of
the scaled noncatalized self-replication ratio for (bottom to top)
$\tilde{c} = 0,1,2,3$ and $4$. The initial concentrations
are (a) $\rho_0 = 1$ and (b) $\rho_0 = .001$.}
\label{fig:rho_mf}
\vspace{0.5cm}
\end{figure}

\begin{figure}[tbp]
\vspace{6.5cm}
\includegraphics{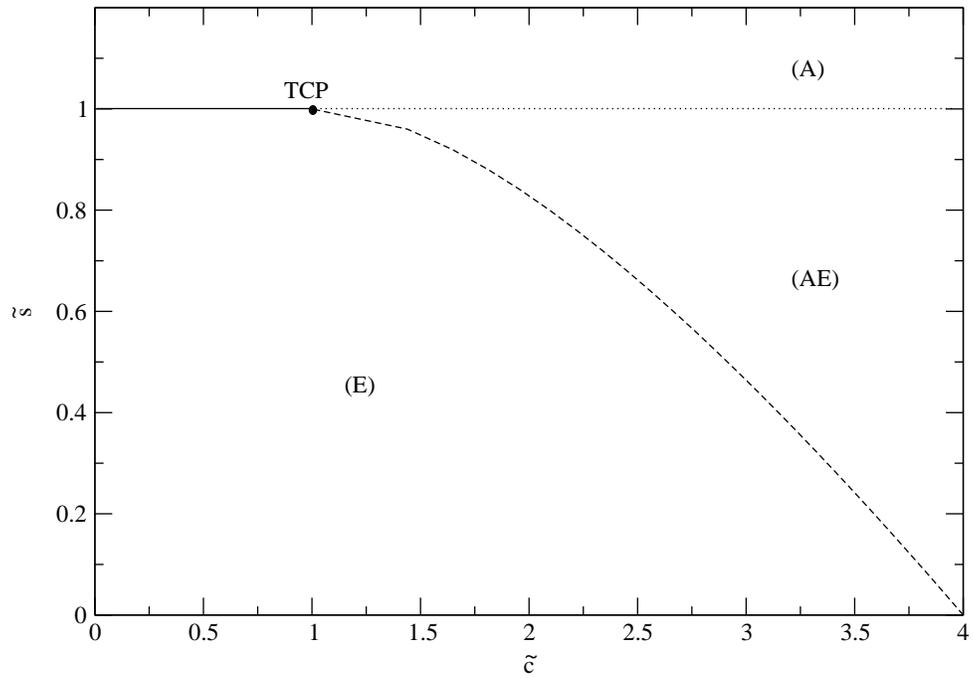} 
\vspace{1cm}
\caption{Mean-field phase diagram in the plane $(\tilde{c},\tilde{s})$
showing the regions of stability of the different steady-state solutions.
The continuous transition ending at the tricritical point (TCP)
is represented as a solid line and the  broken lines indicate
discontinuous phase transitions.}
\label{fig:diag_mf}
\vspace{0.5cm}
\end{figure}

The mean-field limit describes exactly an infinite population
of reactants in an ideally mixed medium
and so it is equivalent to the usual chemical kinetics formulation.
Neglecting spatial correlations among cells, i.e., assuming that at any time
the molecules are distributed randomly  over the lattice cells, it is
straightforward to write the evolution equation for the density of 
replicators or occupied cells at time $t$, namely,
\begin{equation}\label{rho-t}
\rho_{t+1} = \rho_t \left ( 1- \gamma \right ) + 4 \rho_t \left ( 1-\rho_t \right ) 
\left ( s + 4 c \rho_t \right ) .
\end{equation}
We will consider only the stationary solutions $\rho_{t+1}=\rho_{t}=\rho$ of
this equation. The absorbing (empty) state $\rho=0$ is always a solution, while
the non-zero solutions are given by the roots of the quadratic equation
\begin{equation}\label{rho_eq}
\tilde{c} \rho^2 - \rho(\tilde{c} - \tilde{s})+1-\tilde{s} =0 
\end{equation}
where we have introduced the dimensionless parameters 
\begin{equation}\label{tilde}
\tilde{c}  =  \frac{16 c}{\gamma} ~~~\mbox{and}~~~\tilde{s}  =  \frac{4s}{\gamma}.
\end{equation}
This equation has real roots provided that
the condition $(\tilde{c}+\tilde{s})^2 \ge 4 \tilde{c}$ is satisfied. In addition
we can easily show that: (i) for $ \tilde{s} < 1 $ and $\tilde{c} > \tilde{s}$
both roots are positive; (ii) for $ \tilde{s} < 1 $ and $\tilde{c} < \tilde{s}$
both roots are negative; and (iii) for $ \tilde{s} > 1 $ only one of the roots is 
negative. Furthermore, though the solution $\rho = 0$ exists in the entire
plane $(\tilde{c},\tilde{s})$,  it is stable only
if the condition
\begin{equation}
\left .\frac{\partial{\rho_{t+1}}}{\partial{\rho_t}}\right|_{\rho=0}
 = 1 - \gamma + 4 s < 1 ,
\end{equation}
which reduces to $\tilde{s} < 1$, is satisfied. 
In the region where the non-zero roots are physical (i.e., real
and positive),  the stable  root is always the largest one.
In Fig. \ref{fig:rho_mf} we show the steady-state density of
replicators for two different choices of  initial density.
We identify three distinct phases:
the absorbing or empty phase $(E)$ associated to the solution
$\rho=0$; the replicating or active phase $(A)$ associated to the solution
$\rho > 0$; and the phase labeled $(EA)$ where both
solutions $\rho=0$ and $\rho > 0$ are stable. In this phase,
the outcome of the dynamics
is not determined by the control parameters only  but
also by the initial abundance of replicators. 
>From this figure it is clear that the system undergoes a continuous
non-equilibrium phase transition from phase $(E)$ to phase $(A)$
at $\tilde{s} = \tilde{s}_c = 1$ and $\tilde{c} < 1$. Explicitly, near this
transition the density of replicators can be written as
\begin{equation}
\rho \approx \frac{\tilde{s} - 1}{1- \tilde{c}} .
\end{equation}
The continuous transition ends at the tricritical
point $\tilde{s}_t= \tilde{c}_t = 1$ so that for $\tilde{c} > \tilde{c}_t$
the transition between phases $(E)$ and $(EA)$, that takes place at
$\tilde{s} = 2\sqrt{\tilde{c}} - \tilde{c}$,
as well
as the transition between phases $(EA)$ and  $(A)$, that occurs at
$\tilde{s} = 1$, 
are discontinuous. In particular, 
the jumps of the densities of replicators are
\begin{equation}
\Delta \rho = 1 - 1/\tilde{c}^{1/2} ~~~~~~ \tilde{c} >1
\end{equation}
at the transition $(E) - (EA)$, and  
\begin{equation}
\Delta \rho = 1 - 1/\tilde{c} ~~~~~~ \tilde{c} >1
\end{equation}
at the transition $(EA) - (A)$.
These results
are conveniently summarized in the phase diagram shown in Fig.
\ref{fig:diag_mf}. 
It is interesting to note that setting 
$\tilde{s} = \tilde{c}$  we find, close to the tricritical point,
\begin{equation}
\rho \approx \left ( \tilde{s} - 1 \right )^{1/2}
\end{equation}
so that the tricritical point is not in the same universality class 
as the transition observed in the absence of catalytically assisted
replication.

The interpretation of our results within
the prebiotic evolution context
leads to the conclusion 
that for finite values of $\tilde{c}$
an obligatory sexual replicator cannot emerge spontaneously (i.e.,
appear at vanishingly small concentrations). For instance, 
for $\tilde{s}=0$, the minimal initial density of replicators
necessary to engender a prosperous population  is
$\rho=1/2$ at the transition point $\tilde{c}=4$, vanishing
as $1/\tilde{c}$ for large $\tilde{c}$. Actually, an initial colony
of replicators is certain to grow from vanishingly 
small  concentrations provided that  $\tilde{s} > 1$. 
These conclusions, however, must be taken with caution
 since 
in the deterministic limit a vanishingly small concentration 
 means an infinite population of replicators, while
one would expect the first replicators to show up as a single
or  a few copies at most. Of course, a proper understanding of
this emergence phenomenon calls for a stochastic approach, which is the
subject of the next section.
 
%----------------------------------------------------------------
%
\section{The position-fixed limit}\label{sec:level4}
%
%----------------------------------------------------------------

The primary aim of this section is to determine what features of the 
rich phase-diagram obtained in the mean-field limit show up also
in the opposite limit, where the replicators are fixed on the lattice 
cells. 
This is a  rather challenging enterprise as, at least in the case of equilibrium
phase-transitions, there is no totally unambiguous way by which one can detect
the order of the transition through the analysis of finite systems alone \cite{Challa}.
Nevertheless, we tackle this problem using both a steady-state approach for
finite lattices and the dynamical Monte Carlo method for   
lattices of effectively infinite size.

\begin{figure}[tbp]
\vspace{6.5cm}
\includegraphics{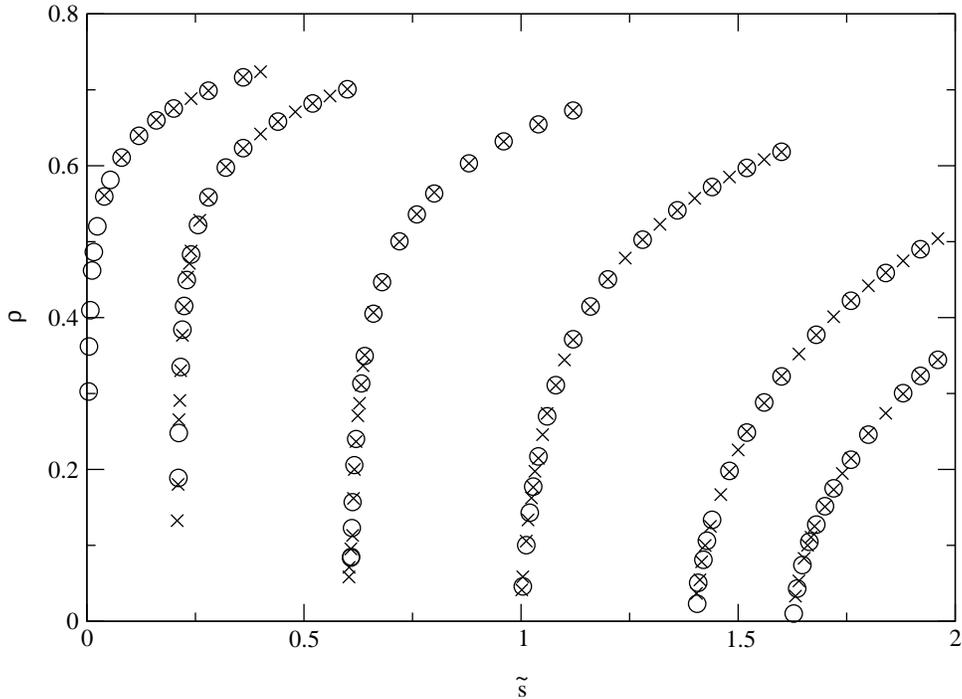} 
\vspace{2cm}
\caption{Average density of replicators $\rho$ as a function of $\tilde{s}$ for 
(left to right) $\tilde{c}=5.70, 5.13, 3.94, 2.58, 0.99$ and
$0$. The lattice sizes are  $L = 100 (\times)$ and $200 (\bigcirc)$.}
\label{fig:rho}
\vspace{1.cm}
\end{figure}

\begin{figure}[tbp]
\vspace{6.5cm}
\includegraphics{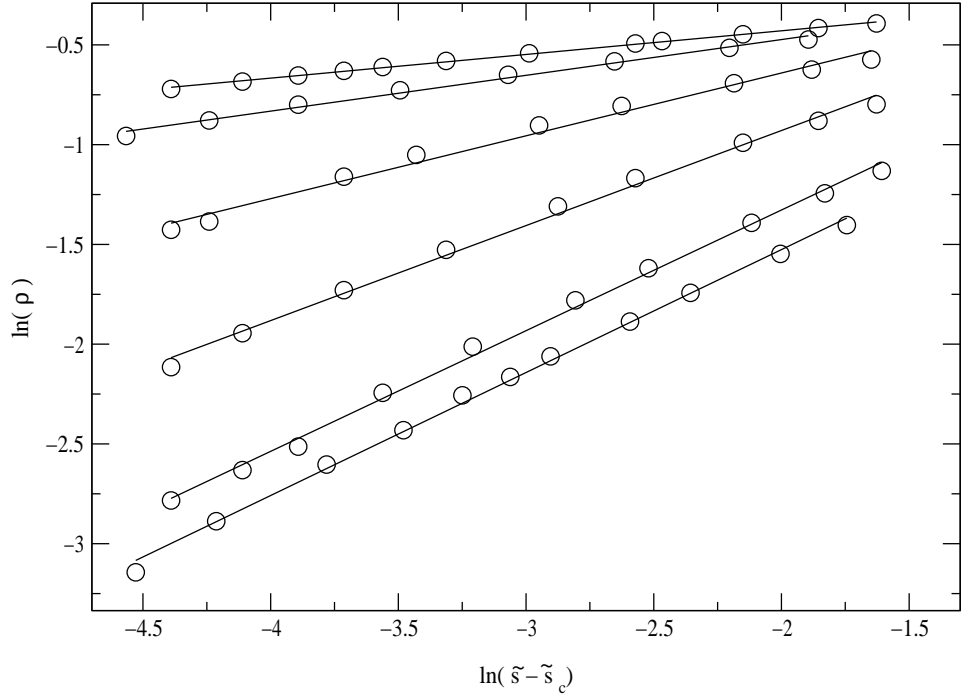} 
\vspace{2cm}
\caption{Logarithm plot of the average density $\rho$ as a function of
$\left ( \tilde{s} - \tilde{s}_c \right )$  for 
(top to bottom) $\tilde{c}= 5.70, 5.13, 3.94, 2.58, 0.99$ and
$0$. The values of $\tilde{s}_c = \tilde{s}_c(c)$ for the different choices 
of $\tilde{c}$ are given in table
\ref{tab:steady} and the straight lines are the numerical fitting obtained
with those data. Only the data for $L=200$ are presented.}
\label{fig:log_rho}
\vspace{1cm}
\end{figure}

First we measure de density $\rho$ of replicators  in
the steady state. Our results for two system sizes ($L=100$ and $L=200$)  are 
shown in Figs. \ref{fig:rho} and \ref{fig:log_rho}.
For each set of the control parameters
we made runs of $10^5$ generations, neglecting the first $2~ 10^4$
generations and recording $\rho$ at  steps of $200$  generations.
A generation corresponds to the simultaneous update of all lattice cells.
Each data point is  the arithmetic mean of these recorded data. Provided that
the population is not extinct, the results are independent  of
the choice of the initial configuration. In particular, we have made runs
starting from all cells occupied or from a seed of only four clustered
occupied cells. Furthermore, as in the mean-field limit we have verified that
our results depend only on the ratios $s/\gamma$
and $c/\gamma$ so throughout the remaining of this paper the value of the
decay constant is held fixed at $\gamma = 0.05$. 
These results clearly indicate the existence of a phase transition
separating the active ($\rho > 0$) and empty ($\rho = 0$) phases of
the model.
Assuming that in the neighborhood of the transition points the density
of replicators goes to zero as $\rho \sim \left ( \tilde{s}- \tilde{s}_c \right )^\beta$
and using the least-square method we can estimate both the critical replication
rate $\tilde{s}_c$ and the critical exponent $\beta > 0$. These estimates are presented
in table \ref{tab:steady} and the quality of the fitting can be appreciated from
Fig. \ref{fig:log_rho}. The statistical errors are of order of $10^{-3}$ but
the  systematic errors, which are due mainly to the difficulty to carry out
long runs very close to the transition point, are probably much larger.   
We note that our estimates of $\beta$ for small values
of $\tilde{c}$ indicate that in this regime
the replicator model belongs to the so-called $(2+1)$
directed-percolation universality class for which $\beta = 0.59 \pm 0.02$
\cite{Essam}.
It should be pointed out that in finite systems the
active regime is a meta-stable one as there is always a finite probability that
the colony becomes extinct due to fluctuations in the stochastic dynamics. Since this
extinction probability increases towards $1$ as the critical point 
is approached, it is very difficult to obtain reliable estimates of $\tilde{s}_c$ and
$\beta$ by means of numerical simulations in the steady-state regime. Of course
we are aware that if the transition happens to become discontinuous at some value
of $\tilde{c} > 0$ then the assumption of a power-law singularity  at the transition point 
breaks down (actually $\beta=0$ in this case).  In fact, the (anomalous) continuous
decrease of $\beta$ as $\tilde{c}$ increases (see table \ref{tab:steady})
is an indication  that this might be the case.
Our careless use of the power-law assumption
is intentional and aims to illustrate the difficulty of detecting the order
of a phase transition using results of steady-state simulations: the 
abrupt variation of $\rho$ at criticality observed in Fig. \ref{fig:rho}
for certain values of $\tilde{c}$, which might indicate the occurrence
of a first-order irreversible transition, can be explained  as 
a continuous transition with a small exponent $\beta$ as well. 

%%%
\begin{table}
\begin{center}
\begin{tabular}{ccc} \hline %kill
$\tilde{c}$ & $\tilde{s}_c$ & $ \beta$ \\ [0.5ex] \hline %\tableline 
0     &1.625  &0.61 \\ 
0.992  &1.400  &0.60 \\ 
2.576  &1.004  &0.47 \\ 
3.936  &0.608  &0.30 \\ 
5.128  &0.210  &0.18 \\ 
5.7008 &0.004  &0.12 \\ \hline %kill
\end{tabular}
\end{center}
\caption{Estimates of the critical point $\tilde{s}_c$ and the
critical exponent $\beta$ assuming the power-law singularity
$\rho \sim (\tilde{s} - \tilde{s}_c )^\beta$.} 
\label{tab:steady}
\end{table}
%%%%
%

We turn now to the analysis of the spreading behavior of a small colony
of replicators settled initially in the center of an otherwise empty lattice
of infinite size. More pointedly, the initial colony 
is composed of four
replicators located in the von Neumann neighborhod of the central
empty cell. Finite size effects
are absent because the lattice size is taken large enough so that during
the time we follow the evolution of the colony  
the replicators can never reach the lattice boundaries. 
This of course sets an upper limit to the number of generations
we can follow the colony and so, in particular, 
we let the population evolve  up to typically $t = 10^4$. As
usual, we  concentrate on the time dependence of the  following key
quantities \cite{Grass-AP}: (i) the average number of replicators $n(t)$; (ii) the
survival probability of the colony $P(t)$; and
(iii) the average mean-square distance over which the replicators have spread
$R^2(t)$. 
For each time $t$ we carry out $M= 2~10^5$ independent runs, all starting with
the same initial colony. Hence $P(t)$ is simply the fraction of runs for 
which there is at least one replicator in the lattice at time $t$. 
Since
$n(t)$ is an average taken over all runs including those which
have already been extinct at generation $t$,
the average number of replicators per surviving
run is given by the ratio $N (t) =n(t)/P(t)$. Furthermore, noting that   
$R^2(t)$ is averaged only over the surviving runs, we can define the fractal
dimension $d_f$ of the surviving colonies of replicators at a given time $t$
as $N \sim R^{d_f}$.

At the transition points we expect that the
measured quantities obey the following scaling laws \cite{Grass-AP}
\begin{eqnarray}
n \left ( t \right ) & \sim  & t^{\eta}  \label{eta}\\ 
P \left ( t \right ) & \sim  & t^{-\delta} \label{delta} \\
R^2 \left ( t \right ) & \sim  & t^{z} \label{z}
\end{eqnarray}
where $\delta$, $\eta$ and $z$ are dynamic exponents which are related
with the fractal dimension of the clusters of replicators through
the equation
\begin{equation}\label{df}
d_f = 2 \frac{\eta + \delta}{z} .
\end{equation} 
In principle, these scaling laws
are valid for continuous as well as discontinuous phase transitions, though the
scaling relations between the exponents, such as the `hyperscaling' relation
\cite{Grass-JPA}
\begin{equation}\label{hyperscaling}
\frac{1}{2}dz - \eta = 2 \delta
\end{equation} 
where $d$ is the lattice dimension, 
hold only in the case of continuous transitions. 

\begin{figure}[tbp]
\vspace{6.5cm}
\includegraphics{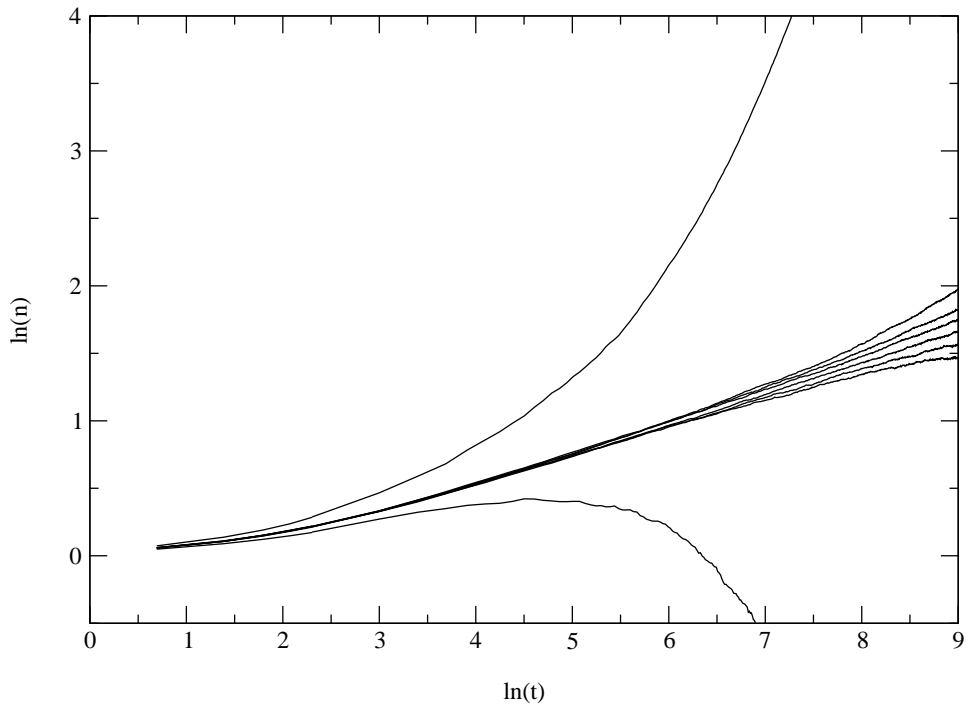} 
\vspace{2cm}
\caption{The log-log plot of $n(t)$ as a function of $t$
for $\tilde{c}=0$ and (top to bottom) $\tilde{s} =1.80, 1.632, 1.630, 1.629, 
1.628, 1.627, 1.626$, and $1.52$. }
\label{fig:logn_s}
\vspace{1.cm}
\end{figure}

\begin{figure}[tbp]
\vspace{6.5cm}
\includegraphics{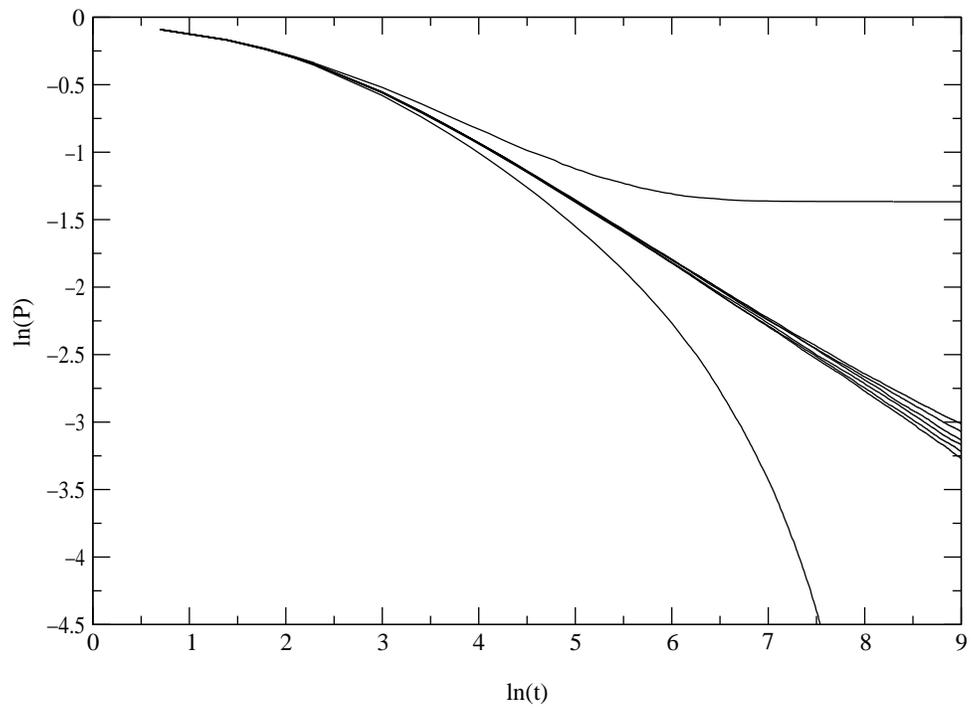} 
\vspace{2cm}
\caption{Same as fig. \ref{fig:logn_s} but for $P(t)$.}
\label{fig:logP_s}
\vspace{1.cm}
\end{figure}

In Figs. \ref{fig:logn_s} and \ref{fig:logP_s}
we present log-log plots of $n(t)$ and $P(t)$, respectively,
as  functions of $t$ in the vicinity of the critical point for $\tilde{c}=0$. 
The dependence of
$R^2(t)$ on $t$ is not shown since, near criticality, the  curves for different
values of $\tilde{s}$ are clustered 
together and do not reveal any qualitatively relevant 
information on the colony  evolution. 
The asymptotic straight lines observed in these figures 
are the signature of critical behavior while upward and downward deviations 
indicate supercritical and subcritical behaviors, respectively. 
A precise estimate for the
critical exponents is obtained by considering the local slopes of the
curves shown in the previous figures. For instance, 
the local slope $\delta (t)$ is defined by  \cite{Grass-JPA,Iwan}
\begin{equation}
-\delta \left ( t \right ) = \frac{\ln \left[ P(t)/P(t/5) \right]}{\ln 5} ,
\end{equation}
which for large $t$ behaves as
\begin{equation}
\delta \left ( t \right ) \sim \delta + \frac{a}{t}
\end{equation}
where $a$ is a constant. 
Analogous expressions hold for $\eta(t)$ and $z(t)$.
Hence plots of the local slopes as functions of $1/t$ allow the calculation
of the critical exponents. Using this procedure we find 
$\tilde{s}_c = 1.628 \pm 0.001 $ which yields 
the exponents $\eta = 0.23 \pm 0.01 $, $\delta = 0.45 \pm 0.01 $ and $z = 1.13 \pm 0.01 $.
The error in $\tilde{s}_c$ is estimated by
determining two values of $\tilde{s}$ as  close as possible to the critical
point for which upward and downward deviations can be observed, while
the errors in the critical exponents are, as usual, the statistical errors obtained
by fitting the local slopes by  straigh lines in the large $t$ regime.
Our exponents are in good agreement with those
of the $(2+1)$ directed-percolation \cite{Grass-JPA} and satisfy very well the
hyperscaling relation (\ref{hyperscaling}) thus indicating
that the transition in the limit $\tilde{c}=0$  is continuous, as expected. 

\begin{figure}[tbp]
\vspace{6.5cm}
\includegraphics{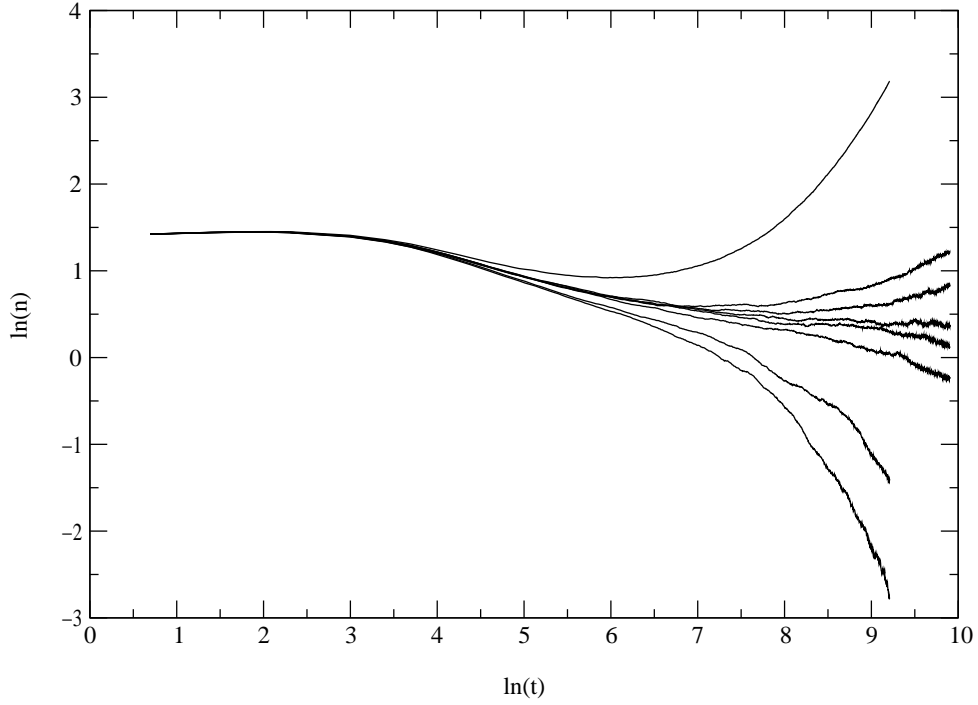} 
\vspace{2cm}
\caption{The log-log plot of $n(t)$ as a function of $t$
for $\tilde{s}=0$ and (top to  bottom) $\tilde{c} = 5.760, 5.712, 5.709, 5.704, 5.702, 
5.701, 5.680$ and  $5.664$.}
\label{fig:logn_c}
\vspace{1.cm}
\end{figure}

\begin{figure}[tbp]
\vspace{6.5cm}
\includegraphics{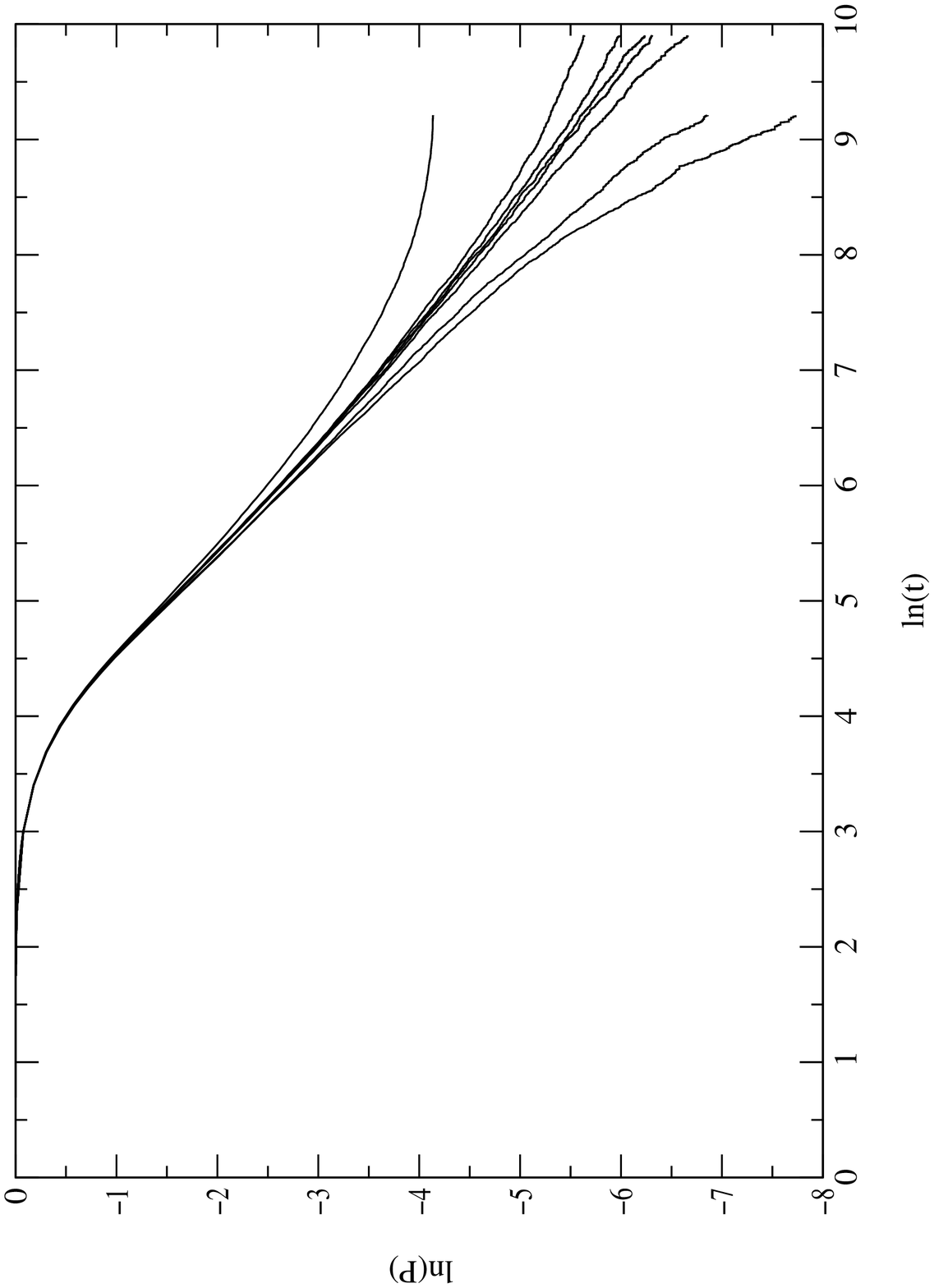} 
\vspace{2cm}
\caption{Same as fig. \ref{fig:logn_c} but for $P(t)$.}
\label{fig:logP_c}
\vspace{0.5cm}
\end{figure}

\begin{figure}[tbp]
\vspace{6.5cm}
\includegraphics{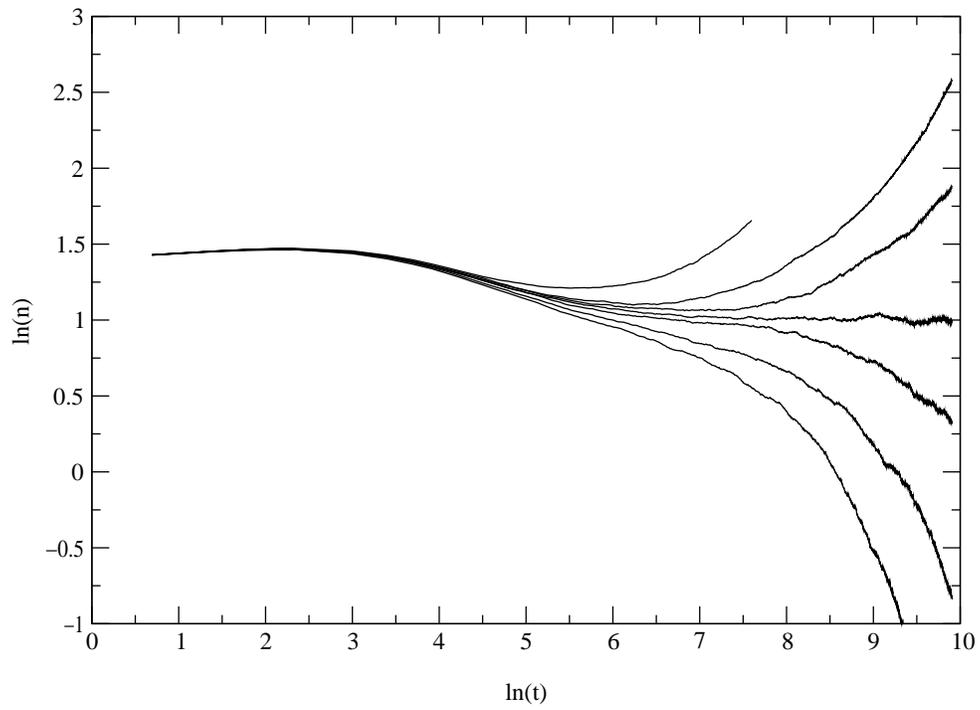} 
\vspace{2cm}
\caption{The log-log plot of $n(t)$ as a function of $t$
for $\tilde{s}=0.2$ and (top to bottom) $\tilde{c} = 5.20, 5.168, 5.160,
5.152, 5.144, 5.136$ and $5.120$. }
\label{fig:logn_0.05}
\vspace{1.cm}
\end{figure}

\begin{figure}[tbp]
\vspace{6.5cm}
\includegraphics{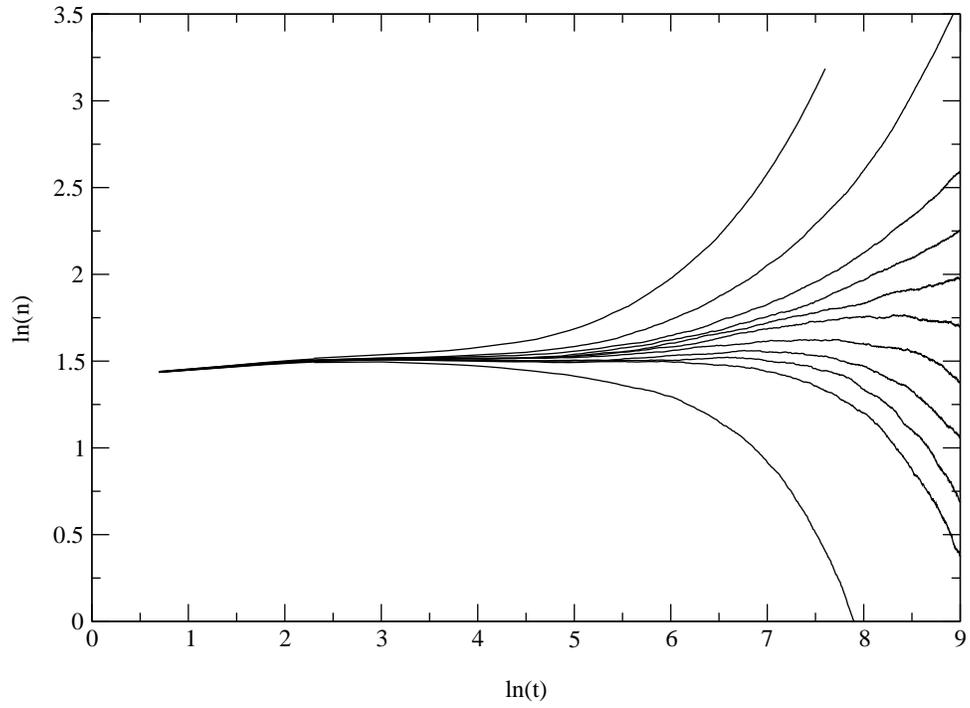} 
\vspace{2cm}
\caption{The log-log plot of $n(t)$ as a function of $t$
for $\tilde{s}=0.6$ and (top to bottom) $\tilde{c} = 4.080, 4.000, 3.968, 3.960, 3.952,
3.944, 3.936, 3.928, 3.920, 3.912$ and $3.840$.}
\label{fig:logn_0.15}
\vspace{1.cm}
\end{figure}

\begin{figure}[tbp]
\vspace{6.5cm}
\includegraphics{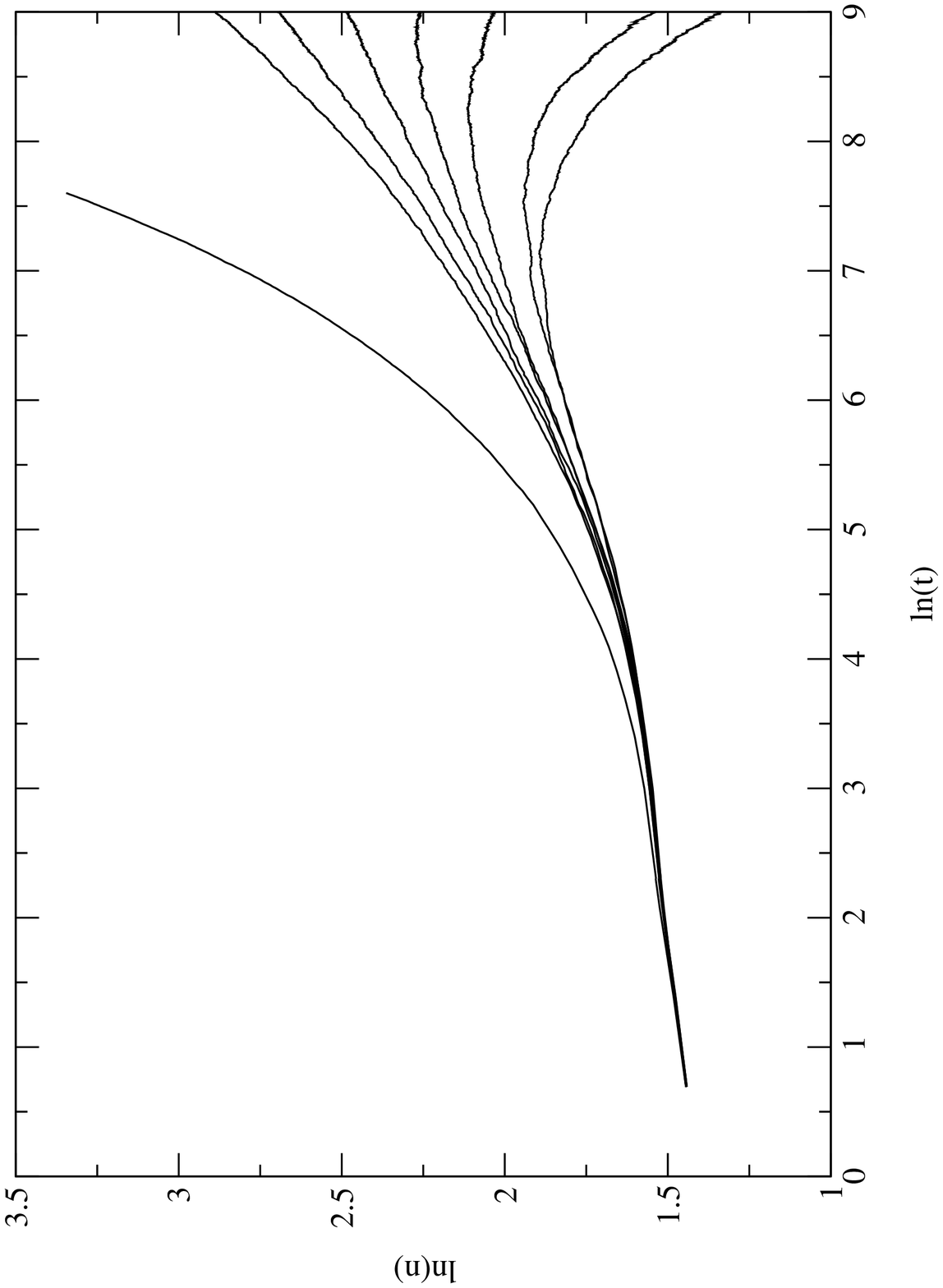} 
\vspace{2cm}
\caption{The log-log plot of $n(t)$ as a function of $t$
for $\tilde{s}=1.0$ and (top to bottom) $\tilde{c} = 2.720, 2.600, 2.592,
2.584, 2.576, 2.568, 2.552$ and $ 2.544$.}
\label{fig:logn_0.25}
\vspace{1.cm}
\end{figure}

The results of the spreading analysis for the other extreme case, $\tilde{s}=0$,
which models a population of obligatory sexual replicators are 
shown in Figs. \ref{fig:logn_c} and \ref{fig:logP_c}. Although 
the dependence of $\ln P(t)$  on $t$ is similar to that
observed in the previous case, the behavior pattern of $\ln n(t)$ 
(see Fig. \ref{fig:logn_c}) is rather
different: in the supercritical regime (i.e.,  $\tilde{c} > \tilde{c}_c$)
$n(t)$ first increases reaching a maximum, then decreases reaching
a minimum and finally  starts to increase monotonically again.
This  pattern is  illustrated better
in Fig. \ref{fig:logn_0.05}, which shows the spreading results for $\tilde{s}=0.2$.
Analysis of these figures, which exceptionally
show  the colony evolution up to $2~10^4$ generations, suggests that
a flat line separates
the supercritical and the subcritical regimes implying thus 
the vanishing of the exponent $\eta$  at the transition point. 
Furthermore, the qualitatively distinct behavior
patterns of $\ln n(t)$ 
illustrated in Figs.  \ref{fig:logn_s}, \ref{fig:logn_c} and  \ref{fig:logn_0.05} 
can be used  to
identify unambiguously the order of the nonequilibrium phase transition
and hence to estimate the location of the tricritical point. To appreciate 
how the  time dependence of $\ln n(t)$ in the supercritical regime
changes continuously 
from the simple monotonic increase for $\tilde{s}=0$ to the complicated
behavior  described above for $\tilde{c}=0$ 
we present in
Figs. \ref{fig:logn_0.15}  and  \ref{fig:logn_0.25} 
log-log plots of $n(t)$ as a function of $t$ for $\tilde{s}=0.6$ 
and $\tilde{s}=1.0$, respectively. In fact, analysis of
Figs. \ref{fig:logn_c} to \ref{fig:logn_0.25}
suggests that the turning point
between those distinct behavior patterns occurs when
the maximum and the minimum of $n(t)$ coincide, i.e., 
the critical curve $\ln n(t)$ vs. $\ln t$ has an inflection point
(see  Fig. \ref{fig:logn_0.15}). The value of $\tilde{s}$ 
and the corresponding $\tilde{c}_c$ at which this behavior occurs
are then identified as the coordinates of the tricritical point.
Applying this procedure we find
$\tilde{s}_t = 0.60 \pm 0.04 $ and $\tilde{c}_t = 4.00 \pm 0.20 $.



The transition points and  the dynamic exponents obtained via
the scaling laws  (\ref{eta})-(\ref{z}) and via the analysis of the local slopes
are summarized in table \ref{tab:dyn_exp}.
The errors in the estimates of the transition points  are calculated 
as described before for the case $\tilde{c}=0$. Except for that case,
we refrain from presenting
the (statistical) errors in the exponents since the systematic errors
are unusually large, due probably to the crossover behavior among (at least) three
different universality classes. For instance, in the vicinity of the
transition point for $\tilde{s} = 0.6$, analysis of the local slope $\delta(t)$
up to $t=500$ indicates a clear tendency to the asymptotic value $\delta \approx  1$
while for  $t >500$ the tendency suddenly changes towards the asymptotic
value $\delta \approx  0.6$.
A similar phenomenon occurs for $\tilde{s}=1.0$ also: the initial tendency is towards
$\delta \approx  0.6$ and then changes towards $\delta \approx  0.45$ for larger times.
As a result the estimate of the exponents becomes strongly dependent on the
precise location of the transition points, which requires even better
statistics as well as much longer runs. We leave this interesting 
research line which includes, for instance,  the identification
of the universality class of the tricritical point
to a future, more technical  contribution.

\begin{figure}[tbp]
\vspace{6.5cm}
\includegraphics{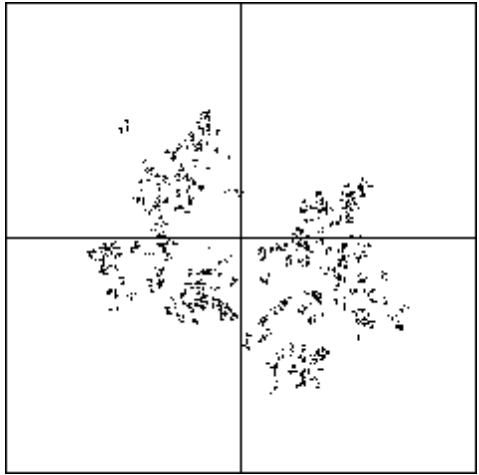} 
\vspace{2cm}
\caption{Snapshot of the lattice configuration at $t=10^4$ showing
the colony of replicators (dots). The parameters are $\tilde{c}=0$
and $\tilde{s} = 1.69$. The initial colony of four replicators was placed in the
middle of the $200 \times 200$ lattice.}
\label{fig:cluster_s}
\vspace{1.cm}
\end{figure}

\begin{figure}[tbp]
\vspace{6.5cm}
\includegraphics{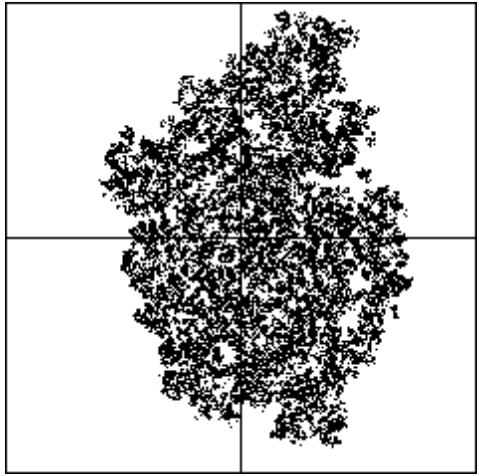} 
\vspace{2cm}
\caption{Same as Fig. \ref{fig:cluster_s} but for $\tilde{s}=0$
and $\tilde{c} =5.92$.}
\label{fig:cluster_c}
\vspace{1.cm}
\end{figure}

%The invariance of the dynamic critical exponents obtained at the various
%second-order transition points, which confirms that the transition
%is in the $(2+1)$ directed-percolation universality class,
%and the closeness
%of the exponents $\delta$ and $z$ to unity at the first-order
%transition points are the worth-mentioning features.
%
\begin{table}
\begin{center}
\begin{tabular} {ccccc} \hline %kill
$\tilde{s}$ & $\tilde{c}_c$ & $ \eta$ & $\delta$ & z\\ [0.5ex] \hline %\tableline
0  & 5.704 $\pm$ 0.005  & -0.03 & 0.96   &  0.98 \\ 
0.2  & 5.152 $\pm$ 0.008 & -0.004  & 0.80  & 1.03 \\ 
0.6  & 3.952 $\pm$ 0.008  & 0.14 & 0.63   & 1.08  \\ 
0.8  & 3.296 $\pm$ 0.008 & 0.22  & 0.54  & 1.11 \\ 
1.0  & 2.584 $\pm$ 0.008 & 0.20  & 0.51  & 1.11 \\ 
1.2  & 1.832 $\pm$ 0.008 & 0.25  & 0.49  & 1.13 \\ 
1.4  & 1.008 $\pm$ 0.008 & 0.23   & 0.47  & 1.11 \\ 
1.628 $\pm$ 0.001 & 0 & 0.23 $\pm$ 0.01  & 0.45 $\pm$ 0.01  & 1.13 $\pm$ 0.01 \\ \hline %kill
\end{tabular}
\end{center}
\caption{Critical dynamic exponents calculated from the slopes of
the straight lines at the transition points $c_c (s)$.} 
\label{tab:dyn_exp}
\end{table}
The evidences in support of our claim that, similarly to the
mean-field limit,  in the position-fixed limit
the nonequilibrium phase transition between the empty
and active phases is discontinuous for small $\tilde{s} $ and so there
is a tricritical point in the phase diagram of the model  are 
threefold:
First, the vanishing of the exponent $\eta$ in this range signalizes  
a distinct asymptotic behavior of the average number of
replicators $n(t)$. Second,  the hyperscaling relation (\ref{hyperscaling}) is clearly
violated for small $\tilde{s}$ while it is satisfied in the regime where we
expect the transition to be a second-order one. Third, 
using Eq. (\ref{df})
and the data of table \ref{tab:dyn_exp}
we find $d_f \approx d = 2$  in the region of  small $\tilde{s}$  indicating that
the clusters of replicators are not fractal objects, 
in contrast to the clusters observed in the vicinity of a second-order transition
(we find, for instance,   $d_f \approx 1.21$ for $\tilde{c}=0$). This point
is illustrated  
in Figs. \ref{fig:cluster_s} and \ref{fig:cluster_c} which show 
snapshots of typical colonies at $t=10^4$ for the two extreme cases.
In both figures the relative distances to the critical points are the same.
The reason for the colonies of obligatory replicators  to be  much denser than
those of malthusian replicators is that the number of replicators
per surviving colony ($N(t) \sim t^{\eta + \delta}$) increases roughly
as $t$ for the former and as $t^{0.7}$  for the latter.
In addition, the 
average square distance over which the initial colony has spread from the center
of the lattice at time $t$  is  roughly of the same order in both extremes
as indicated by the values of the exponent
$z$.


\begin{figure}[tbp]
\vspace{6.5cm}
\includegraphics{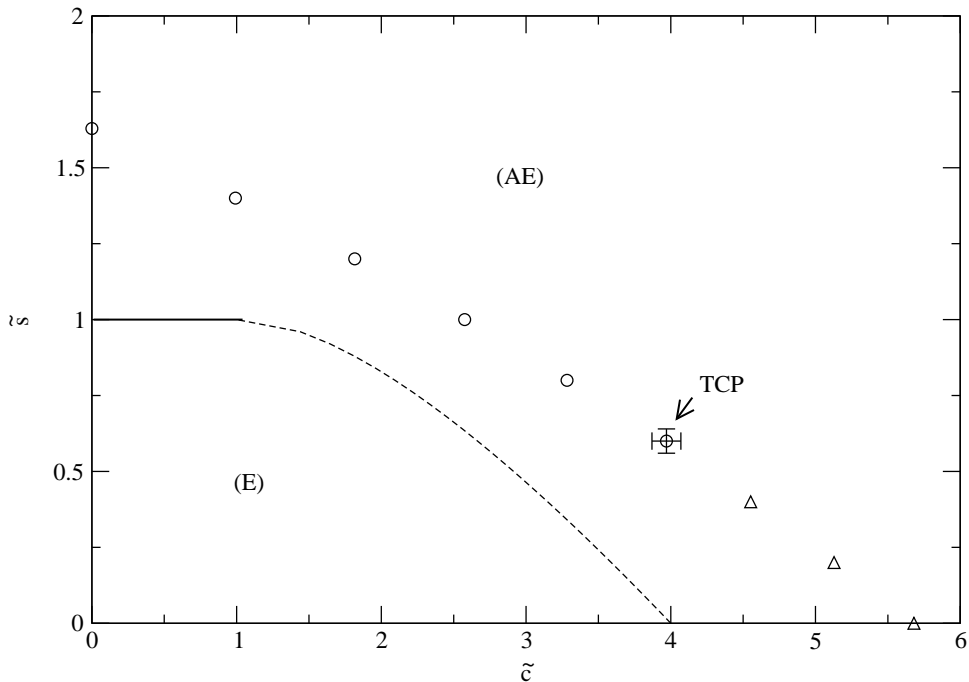} 
\vspace{1cm}
\caption{Phase diagram in the plane $(\tilde{c},\tilde{s})$ for the
position-fixed limit. The continuous and discontinuous transition points
are represented by the symbols $\bigcirc$ and $\bigtriangleup$, respectively.
The error bars represent the uncertainty in the location of the tricritical point
(TCP) while for the other data points the error bars
are smaller than the symbol sizes.
For sake of comparison the mean-field phase-diagram (solid and broken lines)
is also presented.}
\label{fig:diag-fp}
\vspace{0.5cm}
\end{figure}

Finally, in Fig. \ref{fig:diag-fp} we present the phase diagram for
the position-fixed limit. In this case there are only two phases,
namely, the empty phase $(E)$ characterized by a vanishing probability
of survival $P_\infty \equiv \lim_{t \to \infty} P(t) = 0$  
and the phase $(EA)$
where the active and empty states can occur with probabilities
$P_\infty $ and $1-P_\infty$ (see Figs. \ref{fig:logP_s}
and \ref{fig:logP_c}), respectively. 
We note that the population size is effectively
unlimited so that extinction is not certain to occur as in the steady-state
analysis of finite systems.  It is worth
emphasizing that for finite production rates one has  
$P_\infty < 1$ and so there is always a nonvanishing
probability of extinction. (Actually, $P_\infty$ depends on the 
number and location of the replicators in the initial colony,
but the conditions we have chosen are the most relevant for the emergence of
life problem.)
This result contrasts to our findings in the mean-field limit that 
the fixed point associated to the empty phase
becomes unstable for $\tilde{s} > 1$  so that even starting with a vanishingly small
concentration of replicators the population never dies out.
As mentioned before, the reason for this discrepancy is not the
difference in mobility of the replicators but  the fact that
the mean-field analysis actually considers an infinite population and
hence it fails to take into account
the  stochastic fluctuations that could drive a small population to
extinction.

For sake of completeness, we should mention that we have also  carried out a similar
analysis for one-dimensional 
lattices (chains). While the results for the mean-field limit are of course
the same (provided we properly redefine the dimensionless parameters 
$\tilde{c}$ and $\tilde{s}$),
 the fixed-point limit has some distinct features 
that are worth-mentioning.
In particular, we find no  evidence for
a first-order  transition; instead, we find that in both
extremes $\tilde{c}=0$ and $\tilde{s}=0$ the empty and active phases 
are separated by a
second-order phase transition that belongs to the
$(1+1)$ directed-percolation universality class \cite{Grass-AP}. 
Interestingly, in one-dimension the steady-state analysis of finite chains 
is rendered practically useless by the very pronounced finite-size effects,
which are probably due to the proximity to the lower critical dimension of 
the model.

%----------------------------------------------------------------
%
\section{Conclusion}\label{sec:level5}
%
%----------------------------------------------------------------

In this paper we have focussed  on the prior step
in the evolution of life: What are the necessary conditions
for small colonies of molecules
capable of making copies of themselves via some 
template mechanism  to persist? This step must be
passed before one can consider  issues such as
the outcome of the competition between the replicators and their
defective copies \cite{Eigen,Paulo} or between 
different kinds of replicators \cite{hyper,Michod}. 
As a model of replicator we have considered the well-established 
hypercyclic replicator (one-membered hypercycle)
which incorporates two independent mechanisms 
of replication, namely, the direct template replication reaction
(\ref{reac_malt}) and the catalytically-assisted template 
replication reaction (\ref{reac_hyper}), whose rates are
proportional to the parameters $s$
and $c$, respectively.  
Furthermore, motivated by the modern  theories for the evolution of life 
that suggest a
scenario of diffusion-controlled chemical reactions taking place on 
adsorbing surfaces (probably pyrite)  
where each reactant  can move randomly on the surface \cite{Wach,book}, 
we have considered a two-dimensional lattice model where each reactant 
can occupy one of the lattice cells. 

Since the diffusion process of
reactants complicates considerably the analysis, we have
focussed on the two extreme situations:  the infinite diffusion or mean-field limit
and the position-fixed or contact process limit. The expectation is that features
common to both limits should also be present  in the more realistic, intermediate 
situations.  In both cases we  found 
rich phase diagrams  showing the regions  in the plane $(c,s)$ where
the replicators persist (active phase) and die out (empty phase): 
these regions are separated by 
second-order nonequilibrium phase transitions  which turn into
first-order ones at tricritical points. 
The dynamic Monte Carlo method 
has proven very well suited to our investigation 
of the position-fixed limit not only because the method is 
based on  the analysis of the spreading behavior of a small colony of active
cells, which is exactly the  problem we are interested in, but because,
rather surprisingly, it allows an unambiguous identification of the order 
of the nonequilibrium phase transition. 
In addition, we show that the continuous transition is in the universality class 
of the $(2+1)$ directed percolation.

Some remarks on the apparent similarity between Schl\"ogl's first and second
models and the replicator models studied in this paper are in order \cite{Schlogl}. 
In fact, irreversible versions of Schl\"ogl's first and second models are recovered when the 
 reactant $E$ is eliminated from reactions (\ref{reac_malt}) and  (\ref{reac_hyper}), 
respectively, so that the existence of an empty cell containing  source materials  
is not required for replication. Interestingly, this difference is not important
in the case of malthusian replicators ($c=0$) since this model has the
same critical behavior as Schl\"ogl's first model, namely, a second-order
phase transition which is in the same universality class of the
$(d+1)$ directed percolation (the mean-field limit is obtained for $d \geq 4$)
\cite{Janssen}. The comparison between
Schl\"ogl's second model and obligatory sexual replicators ($s=0$) is 
more involved. On the one hand, the mean-field analysis of
Schl\"ogl's second model  predicts
a  first-order phase transition \cite{Schlogl} but Monte Carlo calculations 
indicate that the transition for $d< 4$ is a second-order one which is actually 
in  the same universality class as the transition in Schl\"ogl's first model 
\cite{Grass-ZPB}.
On the other hand, our results for the case $s=0$ show that a
directed-percolation-like, second-order phase transition takes place for
$d=1$ only, the transition being discontinuous for higher dimensions ($d \geq 2$).
We should mention, however, that the Monte Carlo implementation of Schl\"ogl's second model 
actually allows the diffusion of reactants to neighboring cells
and, in addition, allows a cell to shelter more than one reactant \cite{Grass-ZPB} 
so that a comparison with the position-fixed limit may not be appropriate.
In any event, it is our opinion
that the hypercyclic replicator model should not be viewed as a mere 
variant of Schl\"ogl's models; rather it is a well-established model 
of chemical evolution \cite{Eigen,hyper,Michod} which, as far as we know,
has not been studied beyond the mean-field limit.

Although the simple replicator model considered in this paper turned 
out to be a quite exciting  model of nonequilibrium phase transitions,
we should not lose sight of the original purpose of this work and so,
at this stage, 
it is important to highlight the relevance of our results
to the origin of life issue. In fact, the mere existence of a phase
transition between the empty and the active regimes
poses a difficulty to our scenario of the emergence
of life since the scaled production rate $\tilde{s}$ 
of the spontaneously  created 
self-replicating molecule must be larger than some threshold value
already at the outset. 
Though  increasing the mobility of the reactants decreases
 this threshold somewhat, that scenario would be more plausible
if replicators with vanishingly small production rates could also thrive.
The situation becomes even worse in the case of first-order transitions:
in the deterministic mean-field limit the initial abundance of 
the spontaneously created replicators should be large as well,
while in the stochastic position-fixed limit the probability of survival
in the vicinity of  the transition point is some orders of magnitude smaller 
than in the case of  a second-order transition 
(see Figs. \ref{fig:logP_s} and \ref{fig:logP_c}). Furthermore, our results
indicate that some important conclusions, such as the certainty of survival for 
$\tilde{s} > 1$ or the  role played by  the initial concentration
of replicators near a discontinuous transition, are actually artifacts 
of the  deterministic formalism commonly used to study  chemical evolution.

In summary, our results show the necessity of adding some new 
elements to the standard scenario for the emergence of life whose effect
would be to avoid the phase-transitions, allowing thus inefficient replicators
to thrive at this first stage of life. Only then one can invoke
natural selection and imperfect replication to boost the replication rates.
In addition, our results point to the inadequacy of the deterministic 
mean-field or chemical kinetics formulation to  address the
origin of life issue and suggest as an alternative  
stochastic formulation the dynamic Monte Carlo method that has
been extensively used in the physics literature 
to study nonequilibrium phase transitions.

\bigskip

\bigskip

The work of J.F.F. is supported in part by Conselho
Nacional de Desenvolvimento Cient\'{\i}fico e Tecnol\'ogico (CNPq)
and  Funda\c{c}\~ao de Amparo \`a Pesquisa do Estado de S\~ao Paulo 
(FAPESP), Proj.\ No.\ 99/09644-9. C.P.F. 
is supported by FAPESP.

%========================================================================

\end{document}